\documentclass[preprint,showpacs,preprintnumbers,amsmath,amssymb]{revtex4}
\usepackage{graphicx}
\usepackage{dcolumn}
\usepackage{bm}

\font\scripti=cmmi7
\font\scriptscripti=cmmi5
\def\sib#1{\setbox0 = \hbox{\scripti #1}
  \kern-.02em\copy0\kern-\wd0
  \kern.04em\box0} 
\def\ssib#1{\setbox0 = \hbox{\scriptscripti #1}
  \kern-.02em\copy0\kern-\wd0
  \kern.04em\box0} 
\font\tenib=cmmib10 
\skewchar\tenib='177 \skewchar\tenib='177 \skewchar\tenib='177
\def\pbold#1{\setbox0 = \hbox{$ #1 $}
  \kern-.022em\copy0\kern-\wd0
  \kern.011em\copy0\kern-\wd0
  \kern.011em\copy0\kern-\wd0
  \kern.011em\copy0\kern-\wd0
  \kern.011em\box0} 

\usepackage{graphicx}
\usepackage{dcolumn}
\usepackage{bm}

\def\up{\uparrow}
\def\dwn{\downarrow}

\def\lesssim{\ \raise.3ex\hbox{$<$}\kern-0.8em\lower.7ex\hbox{$\sim$}\ }
\def\gesim{\ \raise.3ex\hbox{$>$}\kern-0.8em\lower.7ex\hbox{$\sim$}\ }

\begin{document}
\title{Spin susceptibility and fluctuation corrections in the BCS-BEC crossover regime of an ultracold Fermi gas}

\author{Takashi Kashimura, Ryota Watanabe, and Yoji Ohashi}
\affiliation{Department of Physics, Keio University, 3-14-1 Hiyoshi, Kohoku-ku, Yokohama 223-8522, Japan} 

\date{\today}
\begin{abstract}
We investigate magnetic properties and effects of pairing fluctuations in the BCS (Bardeen-Cooper-Schrieffer)-BEC (Bose-Einstein condensation) crossover regime of an ultracold Fermi gas. Recently, Liu and Hu, and Parish, pointed out that the strong-coupling theory developed by Nozi\`eres and Schmitt-Rink (NSR), which has been extensively used to successfully clarify various physical properties of cold Fermi gases, unphysically gives {\it negative} spin susceptibility in the BCS-BEC crossover region. The same problem is found to also exist in the ordinary non-self-consistent $T$-matrix approximation. In this paper, we clarify that this serious problem comes from incomplete treatment in term of pseudogap phenomena originating from strong pairing fluctuations, as well as effects of spin fluctuations on the spin susceptibility. Including these two key issues, we construct an extended $T$-matrix theory which can overcome this problem. The resulting {\it positive} spin susceptibility agrees well with the recent experiment on a $^6$Li Fermi gas done by Sanner and co-workers. We also apply our theory to a polarized Fermi gas to examine the superfluid phase transition temperature $T_{\rm c}$, as a function of the polarization rate. Since the spin susceptibility is an important physical quantity, especially in singlet Fermi superfluids, our results would be useful in considering how singlet pairs appear above and below $T_{\rm c}$ in the BCS-BEC crossover regime of cold Fermi gases.
\end{abstract}

\maketitle
\par
\section{Introduction} \label{sec1}
\par
The uniform spin susceptibility $\chi$ is a fundamental quantity in considering magnetic properties of an electron system. In a free electron gas, $\chi$ gives useful information about the single-particle density of states at the Fermi level \cite{Landau}. In $s$-wave superconductivity, $\chi$ is suppressed below the superconducting phase transition temperature $T_{\rm c}$ to vanish at $T=0$ \cite{Abrikosov}, because the spin degrees of freedom become inactive by the formation of singlet Cooper pairs. The suppression of the spin susceptibility has been also observed in the underdoped regime of high-$T_{\rm c}$ cuprates, which is referred to as the spin gap phenomenon in the literature \cite{Yasuoka}. Although the origin of the spin gap is still in debate, the importance of preformed pairs has been pointed out \cite{Randeria}.
\par
Since the realization of superfluid $^{40}$K \cite{Regal} and $^6$Li \cite{Zwierlein1,Kinast,Bartenstein} Fermi gases, the high tunability of this quantum system has attracted much attention \cite{Giorgini,Bloch,Chen1}. Indeed, using a tunable pairing interaction associated with a Feshbach resonance \cite{Chin}, one can study superfluid properties from the weak-coupling BCS regime to the strong-coupling BEC limit in a unified manner (BCS-BEC crossover) \cite{Eagles,Leggett,NSR,SadeMelo,Ohashi1}. In the so-called crossover region, a deviation of single-particle excitation spectrum from the free particle dispersion has been observed in the normal state, by using the photoemission-type experiment developed by JILA group \cite{Stewart,Gaebler}. As an explanation for this anomaly, the possibility of the pseudogap phenomenon associated with strong pairing fluctuations has been proposed \cite{Tsuchiya,Chen2,Magierski,Watanabe,Hu,Su,Perali,Mueller}. Since the cold Fermi gas system is much simpler than high-$T_{\rm c}$ cuprates, the former system would be useful for the assessment of the preformed-pair scenario discussed in the latter.
\par
Besides the tunable interaction, the high tunability of population imbalance is another advantage of cold Fermi gases \cite{Zwierlein2,Partridge}. When we describe two atomic hyperfine states in a Fermi gas by pseudospin $\sigma=\uparrow,\downarrow$, a polarized Fermi gas is closely related to an electron system under an external magnetic field. In the limit of low population imbalance, one may evaluate the spin susceptibility. Using this quantity, one can examine whether the preformed singlet pairs really appear in the BCS-BEC crossover regime of a cold Fermi gas. In the case of a finite population imbalance, the mismatch of the Fermi surfaces between the $\uparrow$-spin component and $\downarrow$-spin component is expected to cause the instability of the $s$-wave superfluid state \cite{Clogston}, where various exotic states have been proposed, such as the Fulde-Ferrel-Larkin-Ovchinnikov (FFLO) state \cite{FF,LO} and the Sarma phase \cite{Sarma,LiuWilczek,Sheehy}. 
\par
In this paper, we investigate (pseudo)magnetic properties of a normal state Fermi gas in the BCS-BEC crossover region. In the unpolarized case, the strong-coupling theory developed by Nozi\`eres and Schmitt-Rink (NSR) \cite{NSR} has been extensively used to successfully clarify various physical properties of this system \cite{Ohashi1,Tsuchiya,Chen2,Watanabe,Perali,Pieri,Milstein,OhashiGriffin2,Stajic,Hu2,Diener}. However, when we apply this theory to a polarized Fermi gas, it is known that negative spin susceptibility is obtained in the crossover region \cite{Liu,Parish} (which is thermodynamically forbidden \cite{Sewell}). Because of this serious problem, so far, the phase diagram of a polarized Fermi gas has mainly been examined within the mean-field level \cite{Sheehy}. However, as in the unpolarized case, strong-coupling effects would be also important in the BCS-BEC crossover regime of a polarized Fermi gas. Indeed, it has been pointed out that the FFLO state (which has been predicted in a polarized Fermi gas within the mean-field analysis \cite{Sheehy}) is unstable against pairing fluctuations \cite{Shimahara,Ohashi2}. Thus, to discuss the BCS-BEC crossover physics of a polarized Fermi gas, we need a reliable and tractable strong-coupling theory which can overcome the above mentioned problem. 
\par
In this paper, we show that the ``negative susceptibility problem" also exists in the ordinary (non-self-consistent) $T$-matrix approximation, which has been also extensively used in the unpolarized case. Clarifying the origin of this serious problem, we present a minimal extension of the $T$-matrix theory to correctly give the required positive spin susceptibility in the whole BCS-BEC crossover region. The calculated spin susceptibility in this extended $T$-matrix theory is shown to agree well with the recent experiment on a $^6$Li Fermi gas \cite{Sanner}. We also apply this theory to the system with finite population imbalance, and examine the critical population imbalance at which the superfluid phase transition disappears.
\par
This paper is organized as follows. In Sec.II, we explain our formulation. We also compare our theory with the NSR theory, as well as the ordinary $T$-matrix approximation. In Sec.III, we calculate the spin susceptibility to show that our strong-coupling theory does not meet the negative susceptibility problem in the whole BCS-BEC crossover region. We also compare our results with the recent experiment on a $^6$Li Fermi gas. In Sec.IV, we treat a polarized Fermi gas. Throughout this paper, we set $\hbar=k_{\rm B}=1$, and the system volume $V$ is taken to be unity.
\par
\par
\section{Model polarized Fermi gas and strong-coupling theories} \label{sec2}
\par
We consider a two-component Fermi gas with population imbalance, described by the BCS Hamiltonian,
\begin{equation}
H = \sum_{\bm{p}, \sigma} \xi_{\bm{p}, \sigma} c^\dagger_{\bm{p}, \sigma} c_{\bm{p}, \sigma}
- U  \sum_{\bm{p}, \bm{p}', \bm{q}} c^\dagger_{\bm{p}+\bm{q}/2, \up} c^\dagger_{-\bm{p}+\bm{q}/2, \dwn} c_{-\bm{p}'+\bm{q}/2, \dwn} c_{\bm{p}'+\bm{q}/2, \up}.
\label{H}
\end{equation}
Here, $c^\dagger_{\bm{p}, \sigma}$ is the creation operator of a Fermi atom with momentum $\bm{p}$ and pseudospin $\sigma=\uparrow,\downarrow$, describing two atomic hyperfine states. $\xi_{\bm{p},\sigma} = \varepsilon_{\bm{p}}-\mu_\sigma= \frac{p^2}{2m}-\mu_\sigma$ is the kinetic energy of the $\sigma$-spin component, measured from the Fermi chemical potential $\mu_\sigma$ (where $m$ is an atomic mass). The pairing interaction $-U$ ($<0$) is assumed to be tunable by a Feshbach resonance. As usual, we measure the interaction strength in terms of the $s$-wave scattering length $a_s$, given by
\begin{equation}
\frac{4 \pi a_s}{m} = \frac{-U}{1-U\sum_{\bm{p}}^{\omega_{\rm c}} \frac{1}{2\varepsilon_{\bm{p}}}},
\label{REN}
\end{equation}
where $\omega_{\rm c}$ is a high-energy cutoff. In this scale, the weak-coupling BCS regime and the strong-coupling BEC regime are characterized by $(k_F a_s)^{-1} \lesssim -1$ and $(k_F a_s)^{-1} \gesim 1$, respectively. (Here, $k_F=[3\pi^2 N]^{1/3}$ is the Fermi momentum, where $N$ is the total number of Fermi atoms.) The region $-1 \lesssim (k_F a_s)^{-1} \lesssim 1$ is called the crossover region. In this paper, we consider a uniform Fermi gas, for simplicity.
\par
When we write the chemical potential $\mu_\sigma$ as $\mu_\sigma=\mu+\sigma h$ [where $\mu=(\mu_\uparrow+\mu_\downarrow)/2$ is the averaged chemical potential], Eq. (\ref{H}) may be viewed as a model Hamiltonian for an interacting electron system under an external magnetic field $h$. The spin susceptibility $\chi$ is then given by
\begin{equation}
\chi = \lim_{h\to 0}
{N_\uparrow-N_\downarrow \over h}.
\label{SS}
\end{equation}
Here, $N_\sigma$ is the number of Fermi atoms with $\sigma$-spin, which is calculated from the single-particle thermal Green's function $G_{\bm{p},\sigma} (i\omega_n)$ as
\begin{equation}
N_\sigma = T \sum_{\bm{p},i\omega_n} G_{\bm{p},\sigma} (i\omega_n),
\label{Num}
\end{equation}
where $\omega_n$ is the fermion Matsubara frequency. In this formalism, strong-coupling effects on $\chi$ is described by the self-energy $\Sigma_{\bm{p}, \sigma}(i\omega_n)$ in $G_{\bm{p},\sigma} (i\omega_n)$,
\begin{equation} \begin{split}
G_{\bm{p},\sigma} (i\omega_n) = \frac{1}{\left[ G^0_{\bm{p},\sigma} (i\omega_n) \right]^{-1}-\Sigma_{\bm{p}, \sigma}(i\omega_n)}.
\label{GF}
\end{split} \end{equation}
Here, $G^0_{\bm{p},\sigma} (i\omega_n)= [i\omega_n-\xi_{\bm{p},\sigma}]^{-1}$ is the Green's function for a free Fermi gas. 
\par
\begin{figure}[t]   
\begin{center}
\includegraphics[keepaspectratio,scale=0.75]{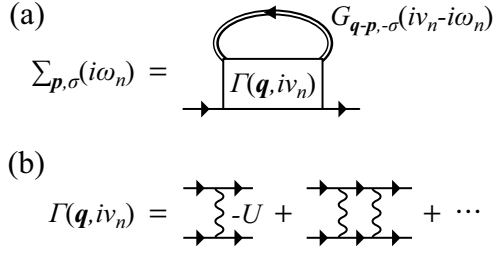}
\caption{(a) Self-energy $\Sigma_{\bm{p}, \sigma} (i\omega_n)$ used in this paper. (b) Particle-particle vertex function $\Gamma (\bm{q}, i\nu_n)$. The solid line and the solid double line represent the free Green's function $G^0$ and the full Green's function $G$ in Eq. (\ref{GF}), respectively. The wavy line describes the attractive interaction $-U$.}
\label{fig1}
\end{center}    
\end{figure}
\par
As mentioned in the introduction, the NSR theory breaks down for a polarized Fermi gas in the sense that it incorrectly gives the negative spin susceptibility ($\chi<0$) in the BCS-BEC crossover region \cite{Liu,Parish}. This implies that one needs to carefully treat the self-energy correction $\Sigma_{\bm{p}, \sigma}(i\omega_n)$ in considering magnetic properties of a polarized Fermi gas. In this paper, we take the strong-coupling corrections diagrammatically described by Fig.\ref{fig1} (We will explain the reason for this choice in Sec.III.), which gives
\begin{equation}
\Sigma_{\bm{p}, \sigma}(i\omega_n) = T \sum_{\bm{q}, i\nu_n} \Gamma (\bm{q},i\nu_n) G_{\bm{q}-\bm{p}, -\sigma} (i\nu_n-i\omega_n),
\label{SE}
\end{equation}
where $\nu_n$ is the boson Matsubara frequency. $\Gamma (\bm{q},i\nu_n)$ is the particle-particle vertex function in the ladder approximation (See Fig.\ref{fig1}(b).),
\begin{equation}
\Gamma (\bm{q}, i\nu_n) = \frac{-U}{1-U \Pi (\bm{q}, i\nu_n)},
\label{VF}
\end{equation}
where
\begin{equation} \begin{split}
\Pi (\bm{q}, i\nu_n) &= T \sum_{\bm{p},i \omega_n} G^0_{\bm{p}+\bm{q}/2,\up} (i\nu_n+i\omega_n) G^0_{-\bm{p}+\bm{q}/2,\dwn} (-i\omega_n) \\
&= - \sum_{\bm{p}} \frac{1-f(\xi_{\bm{p}+\bm{q}/2, \up}) - f(\xi_{-\bm{p}+\bm{q}/2, \dwn})}{i \nu_n-\xi_{\bm{p}+\bm{q}/2, \up} - \xi_{-\bm{p}+\bm{q}/2, \dwn}}
\label{TC}
\end{split} \end{equation}
is the lowest order pair propagator. In Eq. (\ref{TC}), $f(x)$ is the Fermi distribution function.
\par
The ordinary (non-self-consistent) $T$-matrix approximation (TMA) also uses the self-energy in Fig.\ref{fig1}, except that the full Green's function $G$ in Fig.\ref{fig1}(a) is replaced by the noninteracting one $G^0$, as
\begin{equation} 
\Sigma^0_{\bm{p}, \sigma}(i\omega_n) = T \sum_{\bm{q}, i\nu_n} \Gamma (\bm{q},i\nu_n) G^0_{\bm{q}-\bm{p}, -\sigma} (i\nu_n-i\omega_n).
\label{SE0}
\end{equation}
In this sense, our strong-coupling theory may be regarded as an extended $T$-matrix approximation (ETMA) \cite{NOTE1}. We briefly note that, although the NSR theory also uses $\Sigma^0_{\bm{p}, \sigma}(i\omega_n)$, the Green's function in Eq. (\ref{GF}) is expanded to $O(\Sigma^0)$ as
\begin{eqnarray}
G^{\text{NSR}}_{\bm{p},\sigma} (i\omega_n) = G^0_{\bm{p},\sigma} (i\omega_n) + G^0_{\bm{p},\sigma} (i\omega_n) \Sigma^0_{\bm{p}, \sigma}(i\omega_n) G^0_{\bm{p},\sigma} (i\omega_n).
\label{GF_NSR}
\end{eqnarray}
\par
\begin{figure}[t]   
\begin{center}
\includegraphics[keepaspectratio,scale=0.5]{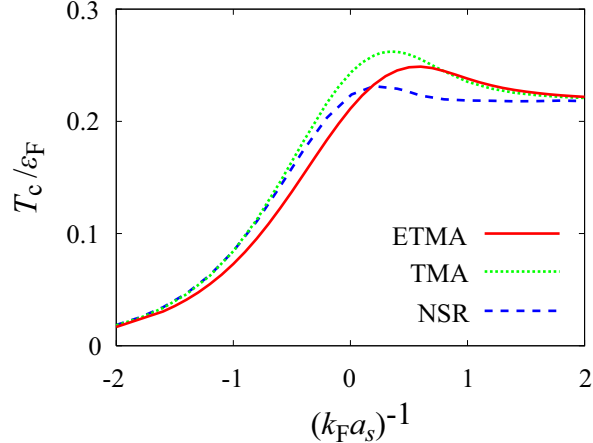}
\caption{(color online) Calculated $T_{\rm c}$ of an unpolarized Fermi gas in the extended $T$-matrix approximation (ETMA). For comparison, we also show the results in the ordinary $T$-matrix approximation (TMA), as well as the NSR theory (NSR). The interaction strength is measured in terms of the inverse scattering length $a_s$, normalized by the Fermi momentum $k_{\rm F}$. $\varepsilon_{\rm F}$ is the Fermi energy.}
\label{fig2}
\end{center}    
\end{figure}

As usual, the superfluid phase transition temperature $T_{\rm c}$ is determined from the Thouless criterion,
\begin{equation}
\Gamma^{-1} (\bm{q}, i\nu_n=0) \big|_{T=T_{\text{c}}}=0.
\label{ThoulessC}
\end{equation}
While the uniform superfluid state corresponds to ${\bm q}=0$, the FFLO state is realized when the highest $T_{\rm c}$ is obtained at $\bm{q} \not= 0$. However, since the latter is known to be unstable against pairing fluctuations even for a weak interaction in the absence of a optical lattice \cite{Shimahara,Ohashi2}, we set $\bm{q}=\bm{0}$ in Eq. (\ref{ThoulessC}) from the beginning. In this case, the (regularized) $T_{\rm c}$-equation is given by
\begin{equation} \begin{split}
\frac{m}{4 \pi a_s} + \sum_{\bm{p}} \left\{ 
\frac{1}
{4\xi_{\bm{p}}} \left[
\tanh{\left( \frac{\xi_{\bm{p}, \up}}{2T} \right)}  + \tanh{\left( \frac{\xi_{\bm{p}, \dwn}}{2T} \right)} \right]
-\frac{1}{2\epsilon_{\bm{p}}}
\right\}
= 0,
\label{ThoulessC2}
\end{split} \end{equation}
where $\xi_{\bm{p}} = \varepsilon_{\bm{p}}-\mu$ is the kinetic energy, measured from the averaged chemical potential $\mu=(\mu_\up+\mu_\dwn)/2$. For a given total number of Fermi atoms $N=N_\uparrow+N_\downarrow$, we solve Eq. (\ref{ThoulessC2}), together with the number equation (\ref{Num}), to determine $T_{\rm c}$, $\mu$, and $h$, self-consistently. In the unpolarized case, the three strong-coupling theories (ETMA, TMA, and NSR) qualitatively give the same BCS-BEC crossover behavior of $T_{\rm c}$, as shown in Fig.\ref{fig2}. In the next section, however, we show that they give very different results for the spin susceptibility.

\begin{figure}[t]   
\begin{center}
\includegraphics[keepaspectratio,scale=0.5]{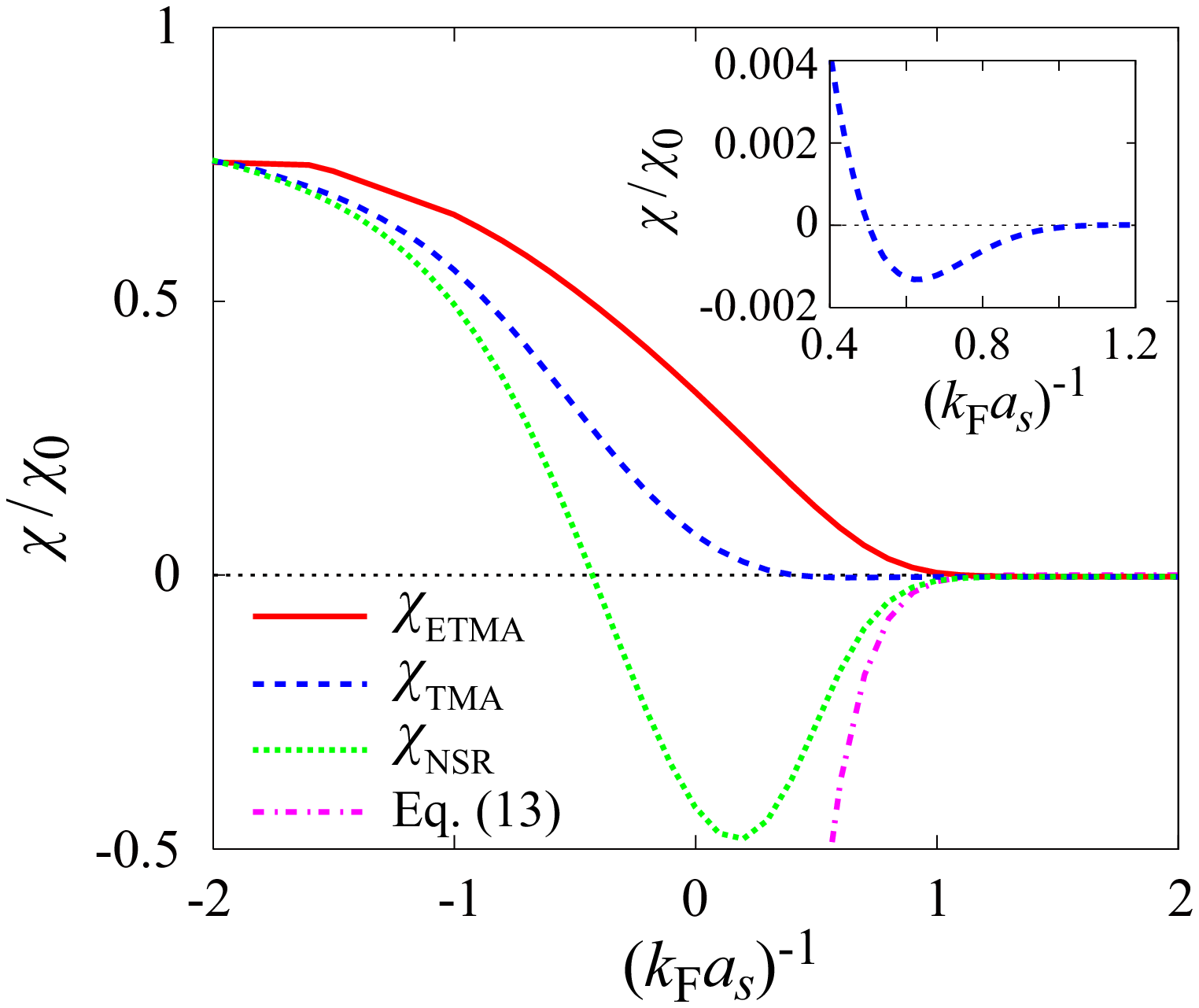}
\caption{(color online) Calculated spin susceptibility $\chi$ at $T_{\rm c}$ in the BCS-BEC crossover. $\chi_{\rm ETMA}$: extended $T$-matrix approximation. $\chi_{\rm TMA}$: ordinary $T$-matrix approximation. $\chi_{\rm NSR}$: NSR theory. The asymptotic form of $\chi_{\rm NSR}$ in Eq. (\ref{AFx}) in the NSR theory is also shown. The inset shows $\chi_{\rm TMA}$ magnified in the crossover region where it becomes negative. $\chi_0$ is the spin susceptibility of a free Fermi gas at $T=0$.}
\label{fig3}
\end{center}    
\end{figure}

\section{Spin susceptibility in the BCS-BEC crossover region}
\par
Figure \ref{fig3} shows the spin susceptibility $\chi$ at $T_{\rm c}$ in the BCS-BEC crossover. As mentioned previously, the NSR theory gives the negative spin susceptibility ($\chi_{\rm NSR}<0$), when the interaction becomes strong to some extent. The situation becomes better in the ordinary $T$-matrix theory ($\chi_{\rm TMA}$). However, as shown in the inset of Fig.\ref{fig3}, $\chi_{\rm TMA}$ slightly becomes negative in the crossover region. In contrast, our extended $T$-matrix approximation ($\chi_{\rm ETMA}$) gives the required {\it positive} spin susceptibility in the whole BCS-BEC crossover. $\chi_{\rm ETMA}$ decreases with increasing the interaction strength, which reflects the increase of preformed Cooper pairs at $T_{\rm c}$. Since all the Fermi atoms form tightly bound singlet molecules in the BEC limit, $\chi_{\rm ETMA}$ vanishes in this limit.
\par
\begin{figure}[t]   
\begin{center}
\includegraphics[keepaspectratio,scale=0.35]{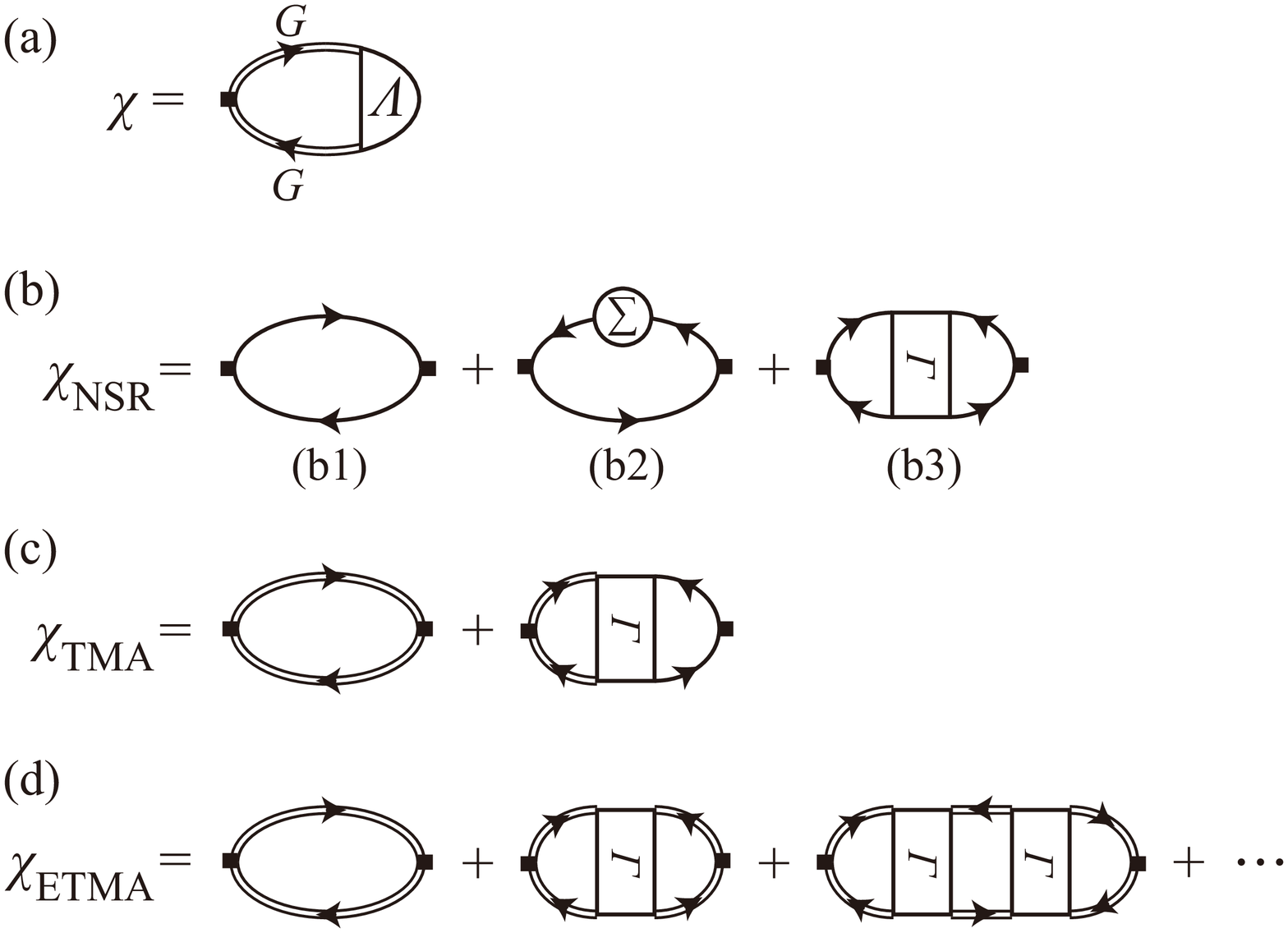}
\caption{(a) Feynman diagram describing spin susceptibility $\chi$. The solid double line is the full Green's function involving the self-energy correction. $\Lambda$ is a three-point vertex part. (b) $\chi_{\rm NSR}$. We only retain the terms to $O(\Sigma^0)$. (c) $\chi_{\rm TMA}$. (d) $\chi_{\rm ETMA}$. In panels (b)-(d), $\Gamma$ is the particle-particle scattering matrix in the ladder approximation in Fig.\ref{fig1}(b).}
\label{fig4}
\end{center}    
\end{figure}
\par
To understand the reason why the present ETMA can overcome the negative susceptibility problem, it is helpful to note that strong-coupling effects on $\chi$ can be divided into the self-energy part and the vertex part, as diagrammatically shown in Fig.\ref{fig4}(a). Between the two, the former comes from the self-energy correction $\Sigma_{\bm{p}, \sigma}(i\omega_n)$ in the single-particle Green's function in Eq. (\ref{GF}), so that this part physically describes how strong-coupling effects on single-particle excitations affect the spin susceptibility $\chi$. In this regard, we recall that strong-pairing fluctuations cause the pseudogap phenomenon in the crossover region \cite{Tsuchiya,Chen2,Magierski,Watanabe,Hu,Su,Perali,Mueller}, where a gap-like structure appears in the normal state density of states $\rho(\omega)$ around the Fermi level $\omega=0$. Since $\chi$ is deeply related to $\rho(0)$ \cite{Landau,NOTE2}, the pseudogap leads to the suppression of $\chi$ in the crossover region.
\par
However, as pointed out in Ref. \cite{Tsuchiya}, the NSR theory overestimates the pseudogap to incorrectly give the {\it negative} density of states around $\omega=0$. This is because of the fact that the NSR theory only retains the self-energy correction to $O(\Sigma^0)$. Thus, the NSR spin susceptibility $\chi_{\rm NSR}$ also becomes negative in the crossover region where the pseudogap becomes remarkable in $\rho(\omega)$. Using the NSR Green's function in Eq. (\ref{GF_NSR}), one finds that $\chi_{\rm NSR}$ is diagrammatically gives by Fig.\ref{fig4}(b). In this panel, the second term ($\equiv\chi_{\rm NSR}^{({\rm b2})}$) describes the pseudogap correction to $\chi$ \cite{Varlamov}, which becomes dominant over the third term (which describes a vertex correction) in the BEC regime. In the BEC limit, one finds
\begin{eqnarray}
\chi_{\text{NSR}}^{\text{BEC}} 
&\simeq& {\bar \chi}_0 + \chi_{\text{NSR}}^{({\rm b2})} 
\nonumber \\
&=& {\bar \chi}_0 -\frac{16 \pi a_s}{m} \left( \frac{2mT_c^{\text{BEC}}}{2\pi} \right)^{\frac{3}{2}}  \zeta \left( \frac{3}{2} \right) \frac{\partial^2 N^0_\up }{\partial h^2 }\Bigg|_{h=0}. 
\label{AFx}
\end{eqnarray}
(We summarize the derivation in Appendix \ref{AppA}.) Here, 
\begin{equation}
\bar{\chi}_0 = {1 \over 2T}\sum_{\bm{p}} 
{\rm sech} ^{2} \left({\xi_{\bm{p}} \over 2T}\right)
\end{equation}
is the spin susceptibility of a non-interacting Fermi gas. In Eq. (\ref{AFx}), $N_\uparrow^0 = \sum_{\bm{p}} f(\xi_{\bm{p}, \uparrow})$ is the number of $\uparrow$-spin atoms in a free Fermi gas. $T_{\rm c}^{\text{BEC}}= 0.218\varepsilon_{\rm F}$ is $T_{\rm c}$ in the BEC limit \cite{NSR,SadeMelo,Ohashi1}. Since the non-interacting part ${\bar \chi_0}$ in Eq. (\ref{AFx}) is remarkably suppressed in the BEC regime due to the negative chemical potential ($\mu<0$) \cite{Leggett,NSR,SadeMelo,Ohashi1}, the correction term $\chi_{\text{NSR}}^{(b2)}$ leads to the  negative spin susceptibility, as shown in Fig. \ref{fig3}. 
\par
The pseudogap effect on $\rho(\omega\sim 0)$ is correctly treated in TMA \cite{Tsuchiya,Chen2,Magierski,Watanabe,Hu,Su,Perali,Mueller}. However, this approximation still has a problem in the vertex part $\Lambda$, so that $\chi_{\rm TMA}$ becomes negative in the crossover region. To see the origin of this, we diagrammatically compare $\chi_{\rm TMA}$ (where the self-energy $\Sigma^0$ in Eq. (\ref{SE0}) is used) with $\chi_{\rm ETMA}$ (where the self-energy $\Sigma$ in Eq. (\ref{SE}) is used) in Fig.\ref{fig4}. While $\chi_{\rm ETMA}$ involves the random phase approximation (RPA)-like series of the Maki-Thompson (MT) diagrams \cite{Varlamov,NOTE3}, TMA only retains this series to the first order. When we approximate the particle-particle scattering matrix $\Gamma$ to the bare interaction $-U$, and ignore all the other interaction effects, $\chi_{\rm ETMA}$ in Fig.\ref{fig4}(d) reduces to the RPA susceptibility,
\begin{equation}
\chi_{\rm ETMA}\simeq 
{{\bar \chi}_0 \over 1+U{\bar \chi}_0}.
\end{equation}
That is, the vertex part $\Lambda_{\rm ETMA}\equiv 1/[1+U{\bar \chi}_0]$, as well as $\chi_{\rm ETMA}$, are always positive. In contrast, because of
\begin{equation}
\chi_{\rm TMA}\simeq 
{\bar \chi}_0[1-U{\bar \chi}_0],
\end{equation}
$\chi_{\rm TMA}$ becomes negative, when the vertex part $\Lambda_{\rm TMA}\equiv 1-U{\bar \chi}_0$ becomes negative \cite{NOTE4}.
\par
Since the present ETMA correctly treats both the self-energy part and the vertex part, the required positive spin susceptibility is obtained over the entire BCS-BEC crossover region, as shown in Fig.\ref{fig3}. 
\par
\begin{figure}[t]   
\begin{center}
\includegraphics[keepaspectratio,scale=0.5]{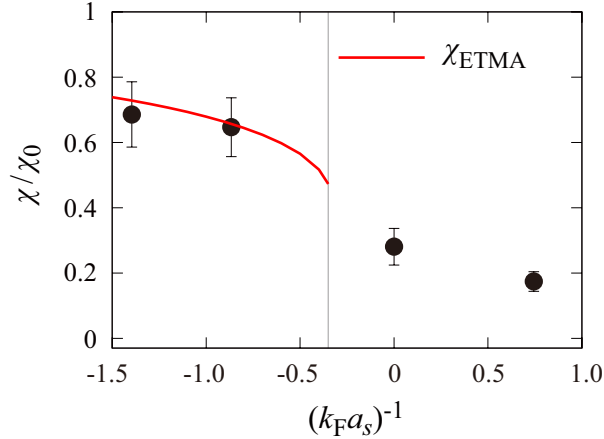}
\caption{(color online) Calculated spin susceptibility $\chi$ in the normal state (solid line). The experimental data \cite{Sanner} are shown as the solid circles. To reproduce the experimental situation \cite{Sanner}, the temperature is fixed at the value of $T_{\rm c}$ for $(k_F a_s)^{-1}=-0.35$ (vertical line). While the left side of the vertical line is the normal phase, the right side is the superfluid phase.
}
\label{fig5}
\end{center}    
\end{figure}

\begin{figure}[t]   
\begin{center}
\includegraphics[keepaspectratio,scale=0.5]{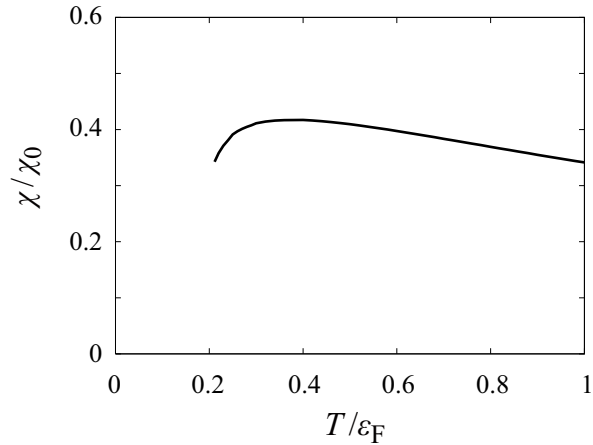}
\caption{Calculated spin susceptibility $\chi$ in the normal state above $T_{\rm c}$. We take $(k_F a_s)^{-1}=0$. Near $T_{\rm c}$, the decrease of $\chi$ with decreasing the temperature is due to the pseudogap effect.}
\label{fig6}
\end{center}    
\end{figure}
\par
In Fig. \ref{fig5}, we compare the calculated spin susceptibility $\chi_{\rm ETMA}$ with the recent experiment on a $^6$Li Fermi gas \cite{Sanner}. In this experiment, the temperature is fixed at the value of $T_{\rm c}$ for $(k_Fa_s)^{-1}=-0.35$, and the spin susceptibility is measured from the {\it in situ} imaging of dispersive speckle patterns. In the normal state above $T_{\rm c}$ (the left side of the vertical line in Fig.\ref{fig5}), $\chi_{\rm ETMA}$ agrees well with the observed spin susceptibility, without introducing any fitting parameter.
\par
While a good agreement with Ref. \cite{Sanner} is obtained, our result is somehow different from the experimental result done by Sommer and co-workers \cite{Sommer}. In this experiment, the observed spin susceptibility in the normal state monotonically increases with decreasing the temperature. In contrast, the calculated spin susceptibility exhibits a peak structure, as shown in Fig. \ref{fig6}. This non-monotonic behavior is similar to the so-called spin gap phenomenon observed in the underdoped regime of high-$T_{\rm c}$ cuprates \cite{Yoshinari}. In the present case, the decrease of $\chi$ near $T_{\rm c}$ is due to the development of the pseudogap in the single-particle density of states.  For this discrepancy between the theory and experiment \cite{Sommer}, although further analyses would be necessary, we note that Refs. \cite{Taylor,Palestini} have recently pointed out that the experimental result may be understood by taking into account the non-equilibrium state associated with a quasi-repulsive interaction.
\par
\begin{figure}[t]   
\begin{center}
\includegraphics[keepaspectratio,scale=0.5]{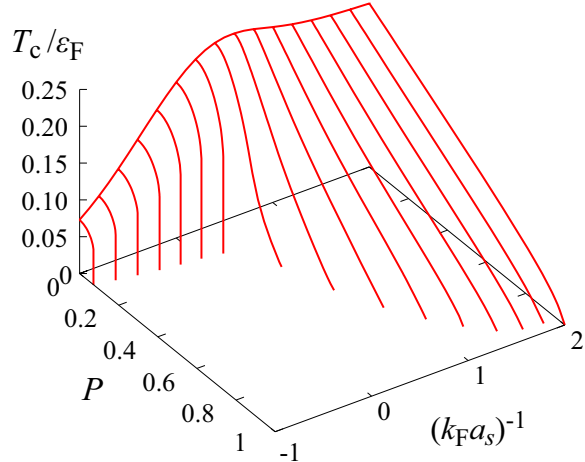}
\caption{(color online) Calculated $T_{\rm c}$, as a function of the interaction strength and the polarization rate $P=[N_\uparrow-N_\downarrow]/[N_\uparrow+N_\downarrow]$. In this figure, we assume the second-order phase transition \cite{NOTE5}.
}
\label{fig7}
\end{center}    
\end{figure}
\par
\section{Polarized Fermi gas in the BCS-BEC crossover regime}
\label{sec4}
\par
We now consider the case of finite population imbalance. Figure \ref{fig7} shows $T_{\rm c}$ in the BCS-BEC crossover regime of a polarized Fermi gas ($N_\uparrow>N_\downarrow$), calculated within the framework of ETMA. We briefly note that, since we are using the Thouless criterion in Eq. (\ref{ThoulessC2}), the second-order phase transition is implicitly assumed. That is, possibility of the phase separation, which is accompanied by the first-order phase transition, is ignored in this figure.
\par
In the strong-coupling BEC limit, the system is well described by a mixture of $N_\downarrow$ tightly bound molecular bosons and $N_\uparrow-N_\downarrow$ excess $\uparrow$-spin atoms. Thus, the superfluid phase transition is dominated by the BEC of the former component. Since the phase transition temperature of an ideal Bose gas is proportional to $N_{\rm B}^{2/3}$ (where $N_{\rm B}$ is the number of bosons), $T_{\rm c}$ in the extreme BEC limit is given by
\begin{equation}
T_{\rm c} = T_{\rm c}^{\text{BEC}}\times 
\left[{N_\uparrow \over (N/2)}\right]^{2/3}
=T_{\rm c}^{\text{BEC}} (1-P)^{2/3},
\label{PdepTc}
\end{equation}
where $T_{\rm c}^{\text{BEC}}=0.218\varepsilon_{\rm F}$ is $T_{\rm c}$ in the BEC limit of a unpolarized Fermi gas. $P=(N_\uparrow-N_\downarrow)/(N_\uparrow+N_\downarrow)$ is the polarization rate. Equation (\ref{PdepTc}) indicates that $T_{\rm c}$ decreases with increasing $P$ to vanish in the fully polarized limit ($P\to 1$).
\par
In the crossover region, as well as the BCS regime, Fig.\ref{fig7} shows that $T_{\rm c}$ vanishes at a certain value of $P$ ($\equiv P_{\rm c}<1$). Since a polarized Fermi gas in the BCS regime is similar to metallic superconductivity under an external magnetic field, the vanishing $T_{\rm c}$ at $P_{\rm c}~(<1)$ is essentially the same as the suppression of the superconducting state by an external magnetic field. In the unitarity limit, one finds $P_{\rm c}=0.13$, which is relatively close to the observed polarization rate $P_{\rm tc}=0.2$ at the tricritical point of a $^6$Li Fermi gas \cite{Shin,NOTE6}.
\par
\begin{figure}[t]   
\begin{center}
\includegraphics[keepaspectratio,scale=0.5]{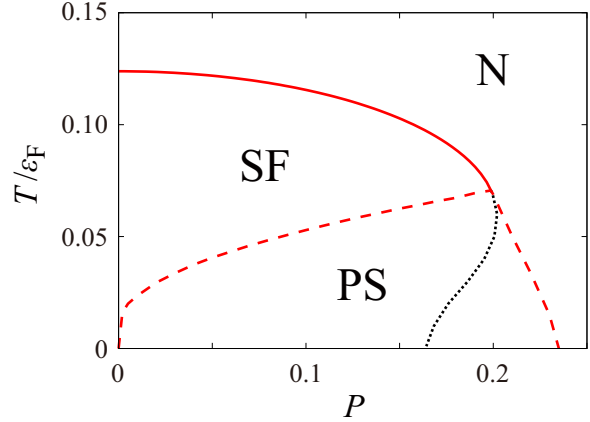}
\caption{(color online) Mean-field phase diagram of a polarized Fermi gas when $(k_F a_s)^{-1}=-1$. The solid line shows the second-order phase transition between the superfluid phase (SF) and the normal state (N). The dashed line shows the first-order phase transition, below which the phase separation (PS) of the superfluid phase and the normal state occurs. The tricritical point is obtained at $P_{\rm tc}=0.199$. When we ignore the PS phase and simply assume the second-order superfluid phase transition, we obtain the dotted line. In this case, one finds $P_{\rm c}=0.202$. Assuming the second-order phase transition (solid line and dotted line), the reentrant region is obtain when $0.164 \le P\le 0.202$ $(=P_{\rm c})$.
}
\label{fig8}
\end{center}    
\end{figure}
\par
In the mean-field theory, the region of the phase separation (PS), which is surrounded by the first-order phase transition line, is obtained in the $T-P$ phase diagram, as shown in Fig.\ref{fig8}. (We summarize how to obtain this figure in Appendix B.) Since the mean-field theory is valid for the weak-coupling regime,  the PS region would also appear in Fig.\ref{fig7}, if one included the possibility of the first-order phase transition beyond the present treatment. To confirm this, however, we need to evaluate the thermodynamic potential $\Omega$, taking into account strong-coupling corrections within the framework of ETMA, which remains as our future problem. 
\par
\begin{figure}[t]   
\begin{center}
\includegraphics[keepaspectratio,scale=0.5]{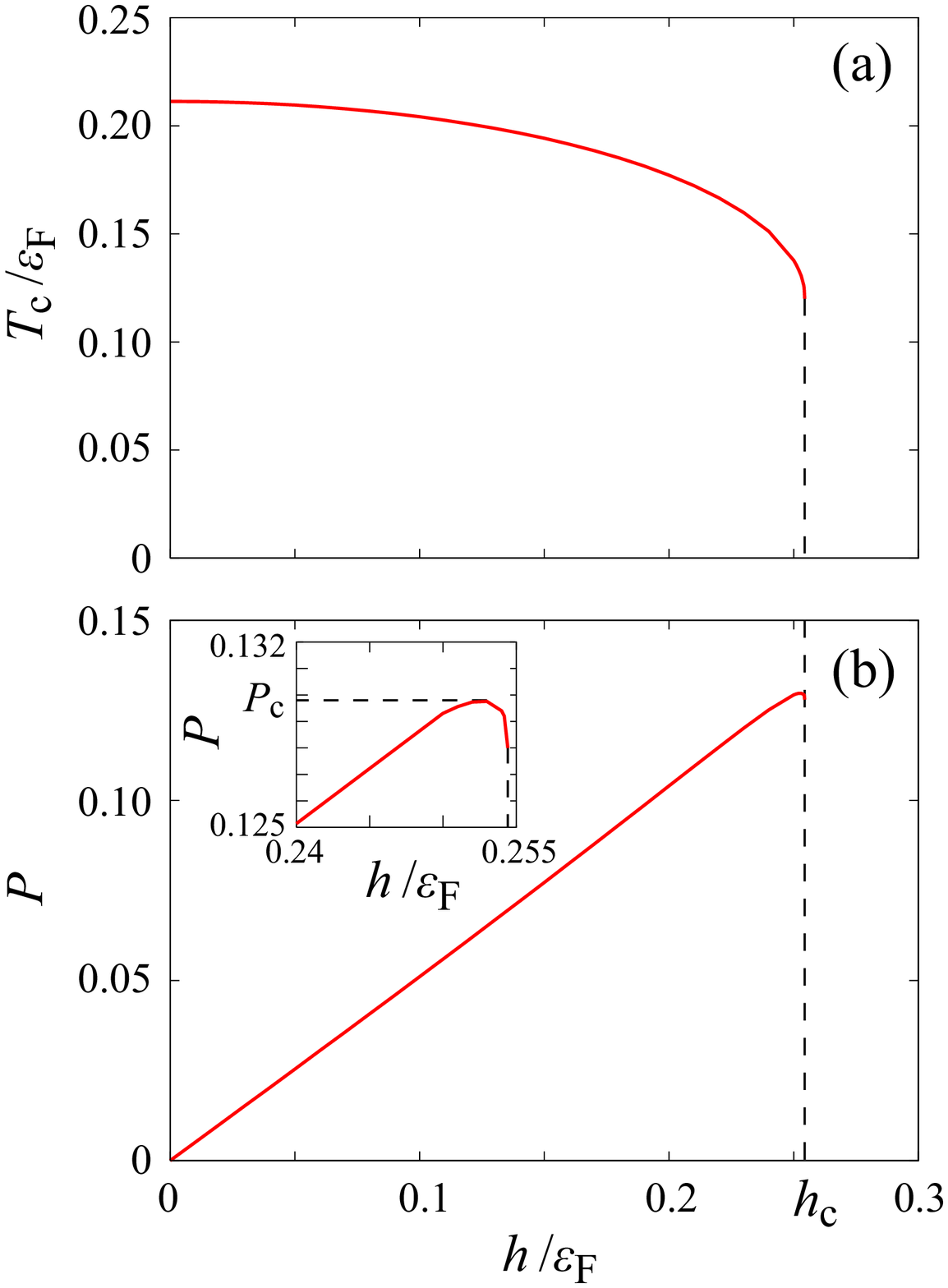}
\caption{(color online) (a) Calculated $T_{\rm c}$, as a function of the ``effective magnetic field" $h$. We take $(k_F a_s)^{-1}=0$. (b) Polarization rate $P$ at $T_{\rm c}$, as a function of $h$. The inset shows the polarization near the critical magnetic field $h_{\rm c}$ at which the second-order phase transition disappears. The critical polarization rate $P_{\rm c}$ is determined as the peak value seen in the inset.
}
\label{fig9}
\end{center}    
\end{figure}
\par
Figure \ref{fig9}(a) shows effects of the ``effective magnetic field" $h$ on the superfluid phase transition. As expected from the magnetic field effect on superconductivity, $T_{\rm c}$ decreases with increasing $h$ to vanish at a critical magnetic field $h_{\rm c}$. When we evaluate the polarization rate $P$ along this $T_{\rm c}$-line, we obtain Fig.\ref{fig9}(b). In this panel, $T_{\text{c}}$ is almost constant around $h=0$, $(dP/dh)_{h\to 0}$ is close to the spin susceptibility $\chi$ $(>0)$. Thus, in a sense, the positive $P$ in panel (b) is a result of the correct treatment of the spin susceptibility in ETMA.
\par
In the inset of Fig.\ref{fig9}(b), one sees a peak structure near the critical magnetic field $h_{\rm c}$. Since $P_{\rm c}$ is given by this peak value, $P_{\rm c}$ is found to obtain, not at $h_{\rm c}$, but below $h_{\rm c}$. As expected from the mean-field phase diagram shown in Fig.\ref{fig8}, one needs a more sophisticated treatment near $P_{\rm c}$ and $h_{\rm c}$ to include the first-order phase transition, as well as phase separation. However, apart from this, the origin of the peak seen in the inset of Fig.\ref{fig9}(b) is explained as follows. When the temperature $T$ is fixed at a certain value, $P$ monotonically increases with increasing $h$. On the other hand, when one decreases the temperature under the condition of a fixed $h$, the polarization $P$ may decrease near $T_{\rm c}$, because of the suppression of the spin susceptibility due to the development of the pseudogap. (See Fig.\ref{fig6}.) In the case of Fig.\ref{fig9}, because the both mechanisms affect $P$, the polarization rate may decrease, when the latter effect becomes dominant. In particular, since the decreases of $T_{\rm c}$ is most remarkable near $h_{\rm c}$ (See Fig.\ref{fig9}(a).), this remarkable decrease of the temperature leads to the decrease of $\chi$, as well as $P$, as shown in the inset of Fig.\ref{fig9}(b). We briefly note that, since the temperature is not fixed in panel (b), the negative value of $(dP/dh)_{h\simeq h_{\rm c}}$ does not mean the violation of the required positivity of the spin susceptibility. In ETMA, the spin susceptibility is always positive, when the temperature is fixed. 
\par
\par
\section{summary} \label{sec5}
\par
To summarize, we have investigated magnetic properties of a cold Fermi gas in the BCS-BEC crossover region. In the NSR theory, it is known that the spin susceptibility becomes negative in the crossover region. We showed that this unphysical result is also obtained in the ordinary (non-self-consistent) $T$-matrix approximation. We clarified that this negative spin susceptibility originates from how to treat the self-energy correction and vertex correction to the spin susceptibility. Improving this, we have succeeded in obtaining the positive spin susceptibility over the entire BCS-BEC crossover region. The calculated spin susceptibility agrees well with the recent experiment done by Sanner and co-workers \cite{Sanner}, without introducing any fitting parameter. We have also applied our extended $T$-matrix theory to a polarized Fermi gas, and have calculated $T_{\rm c}$ in the BCS-BEC crossover region.
\par
In this paper, we have considered the normal state above $T_{\rm c}$. Since the spin susceptibility is known to be strongly suppressed in the superfluid phase below $T_{\rm c}$, it is an interesting next challenge to extend the present theory to include the superfluid order parameter. This extension is also necessary in considering the first-order phase transition, as well as the phase separation, expected in polarized Fermi superfluids. 
\par
In addition, we have only treated a uniform gas, for simplicity. Since a real cold Fermi gas is always trapped in a harmonic potential, the inclusion of this spatial inhomogeneity is crucial for detailed comparison of theoretical results with experimental data. However, since the key issues to overcome the negative susceptibility problem clarified in this paper is also valid for a trapped gas, our results would be useful for the further development of research for magnetic properties of trapped polarized Fermi gases.
\par
\section*{Acknowledgements}
\par
We would like to thank S. Watabe, Y. Endo, D. Inotani, and R. Hanai for useful discussions. Y. O. was supported by Grant-in-Aid for Scientific research from MEXT in Japan (No.22540412, No.23104723, No.23500056).

\appendix
\section{Derivation of Eq. (\ref{AFx})} \label{AppA}
\par
In the NSR theory, strong-coupling correction ($\equiv \chi_{\rm NSR}^{\rm fluct}$) to the spin susceptibility with $O(\Sigma^0)$ is given by 
\begin{equation} 
\label{zms_nsr}
\chi_{\text{NSR}}^{\text{fluc}} = 
{\partial \over \partial h}
\left[
T \sum_{\bm{p},i\omega_n,\sigma} \sigma  G^0_{\bm{p},\sigma} (i\omega_n) \Sigma^0_{\bm{p},\sigma} (i\omega_n) G^0_{\bm{p},\sigma} (i\omega_n)
\right]_{h\to 0}.
\end{equation}
Carrying out the $h$-derivative, we obtain the contributions in Figs.\ref{fig4}(b2) and (b3). Their expressions are given by, respectively,
\begin{eqnarray}
\chi^{({\rm b2})}_{\text{NSR}} &=& -2T \sum_{\bm{p}, i\omega_n, \sigma} \left[ G^0_{\bm{p},\sigma} (i\omega_n) \right]^3 \Sigma^0_{\bm{p},\sigma} (i\omega_n) \Big|_{h=0} \nonumber \\
&=& -T \sum_{\bm{q}, i\nu_n, \sigma} \Gamma (\bm{q}, i\nu_n) \frac{\partial^2}{\partial \mu_\sigma^2} \Pi (\bm{q}, i\nu_n) \Big|_{h=0},\label{zms_nsr1} \\
\chi^{({\rm b3})}_{\text{NSR}} &=& 2T^2 \sum_{\bm{p}, i\omega_n} \sum_{\bm{q}, i\nu_n} \Gamma (\bm{q}, i\nu_n) \left[ G^0_{\bm{p},\up} (i\omega_n) \right]^2 \left[ G^0_{\bm{q}-\bm{p},\dwn} (i\nu_n-i\omega_n) \right]^2 \Big|_{h=0}\nonumber \\
&=& 2T \sum_{\bm{q}, i\nu_n} \Gamma (\bm{q}, i\nu_n) 
\frac{\partial^2}{\partial \mu_\up \partial \mu_\dwn} \Pi (\bm{q}, i\nu_n)\Big|_{h=0}.
\label{zms_nsr2}
\end{eqnarray}
In the BEC limit, the particle-particle vertex function in Eq. (\ref{VF}) reduces to \cite{Haussmann},
\begin{equation}
\Gamma (\bm{q}, i\nu_n) \simeq \frac{8\pi}{m^2 a_s} \frac{1}{i \nu_n - \frac{q^2}{4m}+\mu_{\rm B}}.
\end{equation}
Here, $\mu_{\rm B} = 2\mu + \epsilon_{\rm b}$ may be regarded as the chemical potential of molecular bosons, where $\epsilon_{\rm b} = 1/ma_s^2$ is the binding energy of a two-body bound molecule. Using the fact that the binding energy $\epsilon_{\rm b}$ is very large in the BEC limit ($a_s^{-1}\to \infty$), one may expand $\partial^2 \Pi (\bm{q}, i\nu_n)/\partial \mu_\up^2$ in Eq. (\ref{zms_nsr1}) with respect to $\epsilon_{\rm b}^{-1}$. We then have
\begin{eqnarray}
\frac{\partial^2}{\partial \mu_\up^2} \Pi (\bm{q}, i\nu_n) &=& \sum_{\bm{p}} \frac{1}{i\nu_n-\xi_{\bm{p}+\bm{q}/2,\up} -\xi_{\bm{p}-\bm{q}/2,\dwn}} \frac{\partial^2 f(\xi_{\bm{p}+\bm{q}/2,\up}) }{\partial \mu_\up^2 } + O(\epsilon_{\rm b}^{-2}) \nonumber \\
&\simeq& - \frac{1}{\epsilon_b} \frac{\partial^2 N^0_\up }{\partial \mu_{\uparrow}^2 } + O(\epsilon_{\rm b}^{-2}),
\label{Pi/mu}
\end{eqnarray}
where $N^0_\up = \sum_{\bm{p}} f(\xi_{\bm{p},\up})$ is the number of $\uparrow$-spin atoms in a free Fermi gas. Substituting Eq. (\ref{Pi/mu}) into Eq. (\ref{zms_nsr1}), one obtains
\begin{eqnarray}
\chi^{({\rm b2})}_{\text{NSR}} = 
\frac{2T}{\epsilon_b} \frac{\partial^2 N^0_\up }{\partial h^2 } \Biggr|_{h=0} \sum_{\bm{q}, i\nu_n} \Gamma (\bm{q}, i\nu_n).
\end{eqnarray}
In particular, at $T_{\rm c}$, we find
\begin{equation}
\chi^{({\rm b2})}_{\text{NSR}} = -\frac{16 \pi a_s}{m} \left( \frac{2mT_c}{2\pi} \right)^{\frac{3}{2}}  \zeta \left( \frac{3}{2} \right) \frac{\partial^2 N^0_\up }{\partial h^2 }\Bigg|_{h=0}.
\end{equation}
\par
We briefly note that, because $\partial^2 \Pi (\bm{q}, i\nu_n)/ \partial \mu_\up \partial \mu_\dwn$ is the order of $\epsilon_{\rm b}^{-2}$, one finds $\chi^{({\rm b3})}_{\text{NSR}}=O(\epsilon_{\rm b}^{-2})$. Thus, one can ignore $\chi^{({\rm b3})}_{\text{NSR}}$ in the BEC regime.
\par
\par
\section{Mean-field phase diagram of a polarized Fermi gas} \label{AppB}
\par
In the mean-field theory, the second-order phase transition is determined by solving the ordinary BCS gap equation at $T_{\rm c}$, 
\begin{equation}
1=U\sum_{\bm p}
{1-f(\xi_{{\bm p},\uparrow})-f(\xi_{{\bm p},\downarrow}) \over \xi_{{\bm p},\uparrow}+\xi_{{\bm p},\downarrow}},
\end{equation}
together with the number equation,
\begin{equation}
N=\sum_{{\bm p},\sigma}f(\xi_{{\bm p},\sigma}).
\end{equation}
\par
To evaluate the first-order phase transition temperature, we need to consider the thermodynamic potential $\Omega$ in the presence of phase separation (PS), which is given by
\begin{eqnarray}
\Omega (\mu_\up, \mu_\dwn, T,\Delta, x) = x \Omega_{\text{SF}} (\mu_\up, \mu_\dwn, T,\Delta) + (1-x) \Omega_{\text{N}} (\mu_\up, \mu_\dwn,T).
\end{eqnarray}
Here, $\Omega_{\text{SF}}$ and $\Omega_{\text{N}}$ are the thermodynamic potential in the superfluid (SF) phase and the normal state (N) region, respectively. Their mean-field expressions are given by
\begin{equation} 
\Omega_{\rm SF}= - \frac{m \Delta^2}{4 \pi a_s} + \sum_{\bm{p}} \left[ \xi_{\bm{p}, \downarrow} - E_{\bm{p}, \downarrow} + \frac{\Delta^2}{2\epsilon_{\bm{p}}} \right] -T \sum_{\bm{p},\sigma} \log{ \left( 1+e^{-E_{\bm{p},\sigma}/T} \right)},
\end{equation}
\begin{equation}
\Omega_{\rm N}= -T \sum_{\bm{p},\sigma} \log{ \left(1+e^{-\xi_{\bm{p},\sigma}/T} \right) },
\end{equation}
where $E_{\bm{p}, \sigma}=\sqrt{\xi_{\bm{p}}^2 + \Delta^2}- \sigma h$ is the Bogoliubov excitation energy. Since any intensive variable should have the same value in both the SF region and the N region in the PS phase, each of the chemical potential $\mu_\sigma$ and the temperature $T$ takes the same value in $\Omega_{\text{SF}}$ and $\Omega_{\text{N}}$. The superfluid order parameter $\Delta$ and the volume fraction $x$ of the SF region are, respectively, determined from the stationary conditions of $\Omega$,
\begin{eqnarray}
0 &=& \frac{\partial \Omega}{\partial \Delta} = \frac{\partial \Omega_{\text{SF}}}{\partial \Delta},
\label{GapEq}
\\
0 &=& \frac{\partial \Omega}{\partial x} = \Omega_{\text{SF}} - \Omega_{\text{N}}.
\label{SF=N}
\end{eqnarray}
Equation (\ref{GapEq}) gives the ordinary mean-field BCS gap equation. Equation (\ref{SF=N}) simply means $\Omega_{\rm SF}=\Omega_{\rm N}$. We solve Eqs. (\ref{GapEq}) and (\ref{SF=N}), together with the number equations,
\begin{eqnarray}
N_{\uparrow} &=& x N_{{\uparrow}, \text{SF}} (\mu_\up, \mu_\dwn, \Delta) + (1-x) N_{{\uparrow}, \text{N}} (\mu_\up, \mu_\dwn), 
\label{N1}
\\
N_{\downarrow} &=& x N_{{\downarrow}, \text{SF}} (\mu_\up, \mu_\dwn, \Delta) + (1-x) N_{{\downarrow}, \text{N}} (\mu_\up, \mu_\dwn),
\label{N2}
\end{eqnarray}
to self-consistently determine $\Delta$, $x$, $\mu_\sigma$, below $T_{\rm c}$. In Eqs. (\ref{N1}) and (\ref{N2}), $N_{{\sigma},{\rm SF}}$ and $N_{{\sigma},{\rm N}}$ are the number of $\sigma$-spin atoms in the superfluid region and the normal state region, respectively. 
\par
The phase transition temperature $T_c$ from the PS phase to the normal state is obtained as the temperature at which the superfluid volume fraction $x$ vanishes ($x=0$). The phase boundary between the PS phase and the superfluid phase is determined by the condition $x=1$. We have numerically evaluated these conditions to obtain the phase diagram in Fig. \ref{fig8}. 



\begin{thebibliography}{99}
\bibitem{NOTE1} We note that ETMA still uses the bare Green's functions in the particle-particle scattering vertex $\Gamma$. When the full Green's functions are also used in $\Gamma$, we obtain the self-consistent $T$-matrix theory.
\bibitem{NOTE2} For example, the spin susceptibility $\chi_0$ of a free Fermi gas at $T=0$ is given by $\chi_0 = 2\rho(0)$.
\bibitem{NOTE3} We note that the so-called Aslamazov-Larkin type diagrams vanish identically in the spin susceptibility.
\bibitem{NOTE4} The last term in Fig.\ref{fig4}(b) also has the same physical meaning as $-U{\bar \chi}_0^2$ appearing in TMA.
\bibitem{NOTE5} In the BCS region, as well as the crossover region, one obtains a reentrant behavior of $T_{\rm c}$ near $P_{\rm c}$, where two $T_{\rm c}'$ are obtained for a given $P$. This phenomenon can be also seen in the mean-field phase diagram in Fig.\ref{fig9}, when the second-order phase transition is assumed. However, Fig.\ref{fig9} indicates that the first-order phase transition accompanied by the phase separation cannot ignored near $P_{\rm c}$, so that a more sophisticated treatment would be necessary in determing the phase diagram around the reentrant region near $P_{\rm c}$. In Fig.\ref{fig8}, we only plot the higher $T_{\rm c}$ in the reentrant region, and set $T_{\rm c}=0$ when $P>P_{\rm c}$.
\bibitem{NOTE6} At the tricritical point $P_{\rm tc}$, the superfluid phase transition changes from the second-order type to the first-order type. Although the present calculation ignores the first-order phase transition, the value of $P_{\rm tc}$ is expected to be close to $P_{\rm c}$ determined as the largest value of $P$ where the second-order phase transition occurs. Indeed, in the mean-field phase diagram shown in Fig.\ref{fig8}, one finds that $P_{\rm tc}=0.199$, and $P_{\rm c}=0.202$.
\end{thebibliography}
\end{document}